\documentclass[%
 reprint,
amsmath,amssymb,
 aps,
nofootinbib]{revtex4-2}

\usepackage{graphicx}% Include figure files
\usepackage{braket}
\usepackage{dcolumn}% Align table columns on decimal point
\usepackage{bm}% bold math
\usepackage{xcolor}
\usepackage{slashed}
\usepackage{feynmf} % Load the feynmf package
\usepackage{cancel}
\usepackage{tikz}
\usetikzlibrary{quantikz2}
\begin{document}

%\preprint{APS/123-QED}

\title{Quantum Computational Structure of $SU(N)$ Scattering}% Force line breaks with \\

\author{Navin McGinnis}
 \email{nmcginnis@arizona.edu}
\affiliation{%
 Department of Physics, University of Arizona, Tucson, Arizona 85721, USA}%


%\affiliation{
%\affiliation{
%\affiliation{%


\date{\today}% It is always \today, today,

\begin{abstract}
We study scattering of particles which obey an $SU(N)$ global symmetry through the lens of quantum computation and quantum algorithms. We show that for scattering between particles which transform in the fundamental or anti-fundamental representations, i.e. qudits, all 2-2 scattering amplitudes can be constructed from only three quantum gates. Further, for any $N$, all 2-2 scattering channels are shown to emerge from the span of a $\mathbb{Z}_{2}$ algebra, suggesting that scattering in this context is fundamentally governed by the action of ``bit flips'' on the internal quantum numbers. We frame these findings in terms of quantum algorithms constructed from Linear Combinations of Unitaries and block encoding.
\end{abstract}

\maketitle

%\tableofcontents

%\section{\label{sec:intro}Introduction}
\noindent\textbf{\textit{Introduction}} -
Scattering of particles provides the most direct access to the microscopic laws of nature.  It provides the arena for translating measurements to fundamental interactions described by quantum field theory.  The $S$-matrix encodes this information as a unitary map on the Hilbert space of particle states, representing one of the most complete and general tools for computing observables.  In parallel, quantum computation offers a language for describing microscopic processes as logical operations acting on quantum registers~\cite{Nielsen:2002yeo}.  It is therefore natural to ask whether the $S$-matrix itself possesses a computational structure and, if so, how complex this structure must be for realistic theories of nature.

A naive expectation might be that a theory with a
global $SU(N)$ symmetry requires a large number of
independent operations, at least $N^{2}-1$ for the
generators of the group, in order to parameterize its
scattering amplitudes. In this work we show that this
intuition is misleading. When formulated directly as
quantum operations on the internal
degrees of freedom, every $SU(N)$ invariant 2-2
scattering process between particles in the fundamental
or anti-fundamental representation can be expressed
using only three unitary gates. Each of these operators
turns out to be involutive, and together they generate
$\mathbb{Z}_{2}$ algebras in the two particle channels. The
continuous $SU(N)$ structure of the theory manifests in
the composition of these discrete quantum gates.

This result exposes a remarkable algebraic economy in
the microscopic description of interactions. The $S$-matrix
behaves as a minimal quantum circuit whose logical
structure consists essentially of two ``bit flip''-like
operations and whose closure under composition generates
all $SU(N)$ invariant two to two scattering processes
between qudits. Relations among these gates are fixed
by relations between scattering amplitudes, ensuring that
the resulting processes respect unitarity and analyticity
without requiring any additional dynamical input. In
this sense, the computational structure of scattering is
universal: it is identical for $SU(N)$ for all $N$.

These findings echo recent efforts to understand the
emergence of symmetry from quantum information
resources, such as entanglement and magic. This perspective
has been applied to symmetries in low-energy QCD~\cite{Beane:2018oxh,Low:2021ufv,Beane:2021zvo,Liu:2022grf,Bai:2022hfv,Bai:2023rkc,Bai:2023tey,Liu:2023bnr,Kirchner:2023dvg,Hu:2024hex,Hu:2025lua,Cavallin:2025kjn}, bosonic field theories~\cite{Carena:2023vjc,Kowalska:2024kbs,Chang:2024wrx,Carena:2025wyh,Busoni:2025dns,Gargalionis:2025iqs,Kowalska:2025qmf}, electroweak, flavor, and color interactions in the Standard Model~\cite{Cervera-Lierta:2017tdt,Thaler:2024anb,Liu:2025bgw,Liu:2025qfl,Nunez:2025dch,Nunez:2025xds}, black holes~\cite{Aoude:2020mlg}, strings~\cite{Bhat:2024agd},
suggesting that the algebraic relations that define
symmetries in quantum field theory may themselves arise from the logical
operations permitted within a quantum
system. In tandem, the use of symmetries also find a role in the design and implementation of quantum computation~\cite{Kitaev:1997wr,Gottesman:1997zz,Bacon:2004moa,Harrow:2005pwq,Spekkens:2008acy,Gilyen:2018khw,Faist:2019ahr,Tran:2020azk}. The present work contributes to this program by
showing that the $SU(N)$ structure of two particle
scattering processes reduces to the action of a pair of
involutive unitaries, revealing that the apparent
complexity of a continuous symmetry can be encoded in
a remarkably small set of discrete quantum operations.

Finally, because the 2-2 scattering operators take the form of linear combinations of unitaries, they can be realized using block-encoded quantum circuits with a single ancilla qubit controlling which gate acts on the system.  This provides an explicit computational implementation of scattering amplitudes and a natural bridge between quantum field theory and the circuit-logic language of quantum computation.  The framework developed here provides a concrete framework for exploring how the fundamental interactions of nature can be represented, simulated, or even classified in computational terms.
\noindent\textbf{\textit{Scattering as quantum operations}} - 
To make the
operator viewpoint precise we start from the standard
description of relativistic two-particle scattering, following the notation and conventions in~\cite{McGinnis:2025brt}. Consider particles
of any helicity which additionally carry a finite, \textit{qudit},
degree of freedom described by the states
\begin{equation}
\ket{p,\lambda;i}=\ket{p,\lambda}\otimes\ket{i}.
\end{equation}
We assume that the qudit states have the same dimension for all scattering states and form a complete orthonormal basis of a finite dimensional Hilbert space, $H_{N}=\text{span}\{\ket{1},\ket{2},...,\ket{N}\}$,
\begin{equation}
    \braket{i|j}=\delta_{ij},\quad\sum_{i}\ket{i}\bra{i}=\mathbb{I}_{N}.
    \label{eq:def_rel}
\end{equation}
The scattering states are then normalized in the conventional relativistic conventions
\begin{equation}
    \braket{p^{\prime},\lambda^\prime;j|p,\lambda;i}=(2\pi)^{3}2E_{p}\delta^{(3)}(\vec{p}-\vec{p}^{\prime})\delta_{\lambda\lambda^{\prime}}\delta_{ij}.
\end{equation}

In this article, our main focus will be on scattering amplitudes derived from the Lorentz-invariant $S$-matrix between two-particle asymptotic states
\begin{flalign}\nonumber
    \bra{\{3\},\{4\}}S&\ket{\{1\},\{2\}}=(2\pi)^{4}\delta^{(4)}(p_{1}+p_{2}-p_{3}-p_{4})\\&\times\bra{i_{3}i_{4}}\bra{p_{3},\lambda_{3};p_{4},\lambda_{4}}S\ket{p_{1},\lambda_{1};p_{2},\lambda_{2}}\ket{i_{1}i_{2}},
\end{flalign}
where $\ket{\{i\}}\equiv\ket{p_{i},\lambda_{i};i_{i}}$, and two-particle states are constructed from the direct product $\ket{\{1\},\{2\}}=\ket{p_{1},\lambda_{1}; i_{1}}\otimes\ket{p_{2},\lambda_{2}; i_{2}}$.

The direct product structure between the momentum-helicity degrees of freedom and the qudits allows to decompose $S$-matrix elements as a finite dimensional operator on the bipartite qudit Hilbert space $H_{N}\otimes H_{N}$~\footnote{This can also be argued on the grounds of compactness of kinematic space~\cite{Weinberg:1995mt}.}
\begin{equation}
   S_{kl,ij}(s,\Omega) = \bra{kl}\mathbf{S}(s,\Omega)\ket{ij},
    \label{eq:S_tot}
\end{equation}
where $\mathbf{S}(s,\Omega)=\bra{p_{3},\lambda_{3};p_{4},\lambda_{4}}S\ket{p_{1},\lambda_{1};p_{2},\lambda_{2}}$, defines the matrix elements between momentum and helicity degrees of freedom. For 2-2 scattering, these matrix elements can be parameterized by the center-of-mass energy squared, $s$, and a scattering angle, $\Omega$.

The action of the $S$-matrix as an operator on $H_{N}\otimes H_{N}$ can be decomposed in terms of an operator basis. A complete, linearly independent basis is given by the set $\{\mathbb{I}_{N}\otimes\mathbb{I}_{N},T^{a}\otimes\mathbb{I}_{N},\mathbb{I}_{N}\otimes T^{b},T^{a}\otimes T^{b}\}$, where $T^{a}$ are the generators of the special unitary group, $SU(N)$, satisfying the Lie algebra brackets $[T^{a},T^{b}]=if_{abc}T^{c}$. The normalization has been chosen so that $\text{Tr}(T^{a}T^{b})=\frac{\delta_{ab}}{2}$~\cite{Georgi:1982jb,Ramond:2010zz}. Thus, the action of the $S$-matrix on $H_{N}\otimes H_{N}$ is conveniently parameterized by
\begin{equation}
    \mathbf{S}(s,\Omega)=\mathcal{N}\mathbb{I}_{N}\otimes\mathbb{I}_{N}  + l_a T^{a}\otimes\mathbb{I}_{N} + r_{a}\mathbb{I}_{N}\otimes T^{a} + c_{ab}T^{a}\otimes T^{b},
    \label{eq:S_decomp}
\end{equation}
with the coefficients given by:
\begin{flalign}\nonumber
    \mathcal{N}=&\frac{1}{N^{2}}\text{Tr}(\mathbf{S}(s,\Omega)\mathbb{I}_{N}\otimes\mathbb{I}_{N}),\; l_{a}=\frac{2}{N}\text{Tr}(\mathbf{S}(s,\Omega)T^{a}\otimes\mathbb{I}_{N})\\
    r_{a}=&\frac{2}{N}\text{Tr}( \mathbf{S}(s,\Omega)\mathbb{I}_{N}\otimes T^{a}),\; \frac{c_{ab}}{4}=\text{Tr}( \mathbf{S}(s,\Omega)T^{a}\otimes T^{b}).
\end{flalign}

We further decompose the $S$-matrix elements by subtracting off the disconnected contributions where no scattering occurs
\begin{equation}
    S-1=i\mathcal{T}.
\end{equation}
Thus, the fully-connected components of the 2-2 scattering amplitudes can be similarly realized as a finite dimensional operator on the qudit space
\begin{equation}
    \bra{kl}i\mathbf{T}(s,\Omega)\ket{ij}=(2\pi)^{4}\delta^{(4)}(p_{1}+p_{2}-p_{3}-p_{4})i\mathcal{M}_{kl,ij}(s,\Omega),
\end{equation}
where $\textbf{T}(s,\Omega) = \bra{p_3,\lambda_{3};p_{4},\lambda_4}\mathcal{T}\ket{p_{1},\lambda_{1};p_{2},\lambda_{2}}$. In terms of our choice of basis on $H_{N}\otimes H_{N}$ we have

\begin{equation}
\mathcal{M} = \tilde{\mathcal{M}}\mathbb{I}_{N}\otimes\mathbb{I}_{N}+\mathcal{A}_{a}T^{a}\otimes\mathbb{I}_{N} + \mathcal{B}_{a}\mathbb{I}_{N}\otimes T^{a} + \mathcal{C}_{ab}T^{a}\otimes T^{b},
\label{eq:M_decomp}
\end{equation}
with
\begin{flalign}\nonumber
    \tilde{\mathcal{M}}=&\frac{1}{N^{2}}\text{Tr}(\mathcal{M}\mathbb{I}_{N}\otimes\mathbb{I}_{N}),\;\mathcal{A}_{a}=\frac{2}{N}\text{Tr}(\mathcal{M}T^{a}\otimes\mathbb{I}_{N})\\
    \mathcal{B}_{a}=&\frac{2}{N}\text{Tr}(\mathcal{M}\mathbb{I}_{N}\otimes T^{a}), \; \frac{\mathcal{C}_{ab}}{4}=\text{Tr}(\mathcal{M}T^{a}\otimes T^{b}).
    \label{eq:M_decomp_coeffs}
\end{flalign}
The operator decomposition in Eq.~\ref{eq:M_decomp}\;\&~\ref{eq:M_decomp_coeffs} therefore gives
a complete and finite dimensional description of two
particle scattering amplitudes in the internal space. In particular, it
makes explicit that the $S$-matrix acts as a quantum
operation on the two
qudits. This viewpoint will be central in the next
section, where we impose $SU(N)$ invariance and show
that the entire structure collapses to a remarkably small
set of unitaries.

%%%%%%%%%%%%%%%%%%%%%%%%%%%%%%%%%%%%%%%%%%%%%%%%%%%%%%%%%%%%%%%%%%%%%%%%%%%%%%%%%%%%%%%%%%%%%%%%%%%%%%%%%%%%%%%%%%%%%%%%%%%%%%%%%%%%%%%%%%%%%%%%%%%%%%%%%%%%%%%%%%%%%%%%%%%%%%%%
%\section{Minimal entanglement and conserved charges}
\noindent\textbf{\textit{$SU(N)$ scattering from quantum gates}} - %
Let us specialize the formalism in the previous section to scattering which obeys an $SU(N)$ symmetry. First note that the defining relations of the qudit states are invariant under
\begin{equation}
	\ket{i}^{\prime} = U\ket{i},
\end{equation}
$U\in SU(N)$. Under this action, we may treat the qudit states as transforming according to the fundamental representation $\ket{i}\sim \textbf{N}$, or the anti-fundamental $\ket{i}\sim \bar{\textbf{N}}$ of $SU(N)$. In terms of scattering of two-particle states, amplitudes are realized as quantum operators on the space $\textbf{N}\otimes \textbf{N}$ or $\textbf{N}\otimes \bar{\textbf{N}}$. Under the diagonal action of $SU(N)$, these spaces decompose in terms of the irreducible representations
\begin{equation}
	\textbf{N}\otimes\textbf{N} \simeq \textbf{S}\oplus\textbf{A}
\end{equation}
where $\textbf{S}(\textbf{A})$ is the symmetric (antisymmetric) representation, and
\begin{equation}
	\textbf{N}\otimes\bar{\textbf{N}} \simeq \textbf{1}\oplus\textbf{Adj},
\end{equation}
where $\textbf{1}(\textbf{Adj})$ is the singlet (adjoint) representation.

Assuming that the $S$-matrix is invariant under the global $SU(N)$ symmetry
the scattering amplitudes between two-particles states will act diagonally between irreducible representations and can be concisely decomposed as 
\begin{equation}
	\mathcal{M} = \mathcal{M}_{\textbf{S}}P_{\textbf{S}} + \mathcal{M}_{\textbf{A}}P_{\textbf{A}}
\end{equation}
or
\begin{equation}
	\mathcal{M} = \mathcal{M}_{\textbf{1}}P_{\textbf{1}} + \mathcal{M}_{\textbf{Adj}}P_{\textbf{Adj}},
\end{equation}
depending on whether the scattering occurs between $\textbf{N}\otimes \textbf{N}\to\textbf{N}\otimes \textbf{N}$ or $\textbf{N}\otimes \bar{\textbf{N}}\to \textbf{N}\otimes \bar{\textbf{N}}$, respectively.  
The scalar amplitudes in these relations have been defined by
\begin{equation}
\mathcal{M}_{R} = \bra{kl}\mathcal{M}\ket{ij}(P_{R})_{kl,ij}/\text{Tr}(P_{R}),\end{equation}
where $R = \mathbf{1},\mathbf{Adj}, \mathbf{S},$ or $\mathbf{A}$, and we have also introduced the projection operators
\begin{equation}
(P_{\textbf{S}})_{ij,rs} = \frac{\delta_{ir}\delta_{js} + \delta_{jr}\delta_{is}}{2},\quad\quad (P_{\textbf{A}})_{ij,rs} = \frac{\delta_{ir}\delta_{js} - \delta_{jr}\delta_{is}}{2},
\end{equation}
and 
\begin{equation}
(P_{\textbf{1}})_{ki,pr} = \frac{\delta_{ki}\delta_{pr}}{N},\quad\quad (P_{\textbf{Adj}})_{ki,pr} = \delta_{kp}\delta_{ir} - \frac{\delta_{ki}\delta_{pr}}{N}.
\end{equation}

Note that $P\cdot P = P$ and in terms of the qudit basis, these operators are given by
\begin{flalign}
P_{\textbf{S}} =& \frac{N+1}{2N}\mathbb{I}_{N}\otimes\mathbb{I}_{N} + \sum_{a=1}^{N^{2}-1}T^{a}\otimes T^{a},\\
P_{\textbf{A}} =& \frac{N-1}{2N}\mathbb{I}_{N}\otimes\mathbb{I}_{N} - \sum_{a=1}^{N^{2}-1}T^{a}\otimes T^{a},
\end{flalign}
and 
\begin{flalign}
P_{\textbf{1}} =& \;\mathbb{I}_{N}\otimes\mathbb{I}_{N} - 2\sum_{a=1}^{N^{2}-1}T^{a}\otimes T^{a},\\
P_{\textbf{Adj}} =&  \;2\sum_{a=1}^{N^{2}-1}T^{a}\otimes T^{a}.
\end{flalign}
Although these operators are relatively simple, it is not clear how useful these expansions are for the purposes of quantum simulation and algorithms. Instead let us consider their reducible forms defined by
\begin{flalign}
	P_{\textbf{S}} + P_{\textbf{A}} =& \mathbf{S}_{\mathbb{I}},\\
	P_{\textbf{S}} - P_{\textbf{A}} =& \mathbf{S}_{W},
\end{flalign}
where we find the identity and Swap quantum gates on the $\textbf{N}\otimes \textbf{N}$ space
\begin{flalign}
	\mathbf{S}_{\mathbb{I}} \ket{ij} = \ket{ij},\quad
	 \mathbf{S}_{W}  \ket{ij} = \ket{ji}.
\end{flalign}	
Note that unitarity of these operators is guaranteed by $(P_{\textbf{S}} \pm P_{\textbf{A}})^{\dagger}(P_{\textbf{S}} \pm P_{\textbf{A}}) = P_{\textbf{S}} + P_{\textbf{A}} = \mathbf{S}_{\mathbb{I}}$. On the $\textbf{N}\otimes \bar{\textbf{N}}$ space we have
\begin{flalign}
	P_{\textbf{1}} + P_{\textbf{Adj}} =& \;\mathbf{S}_{\mathbb{I}},\\
	P_{\textbf{1}} - P_{\textbf{Adj}} =& \;\mathbf{U},
\end{flalign}
where $\mathbf{S}_{\mathbb{I}}$ is again the identity. Unitarity is similarly guaranteed as in the previous case. The operator $\mathbf{U}$ given in terms of the qudit basis as 
\begin{equation}
	\mathbf{U} = \left(\frac{2}{N^{2}}-1\right)\mathbb{I}_{N}\otimes\mathbb{I}_{N} - \frac{4}{N}\sum_{a}T^{a}\otimes T^{a},
\end{equation}
detects a charge-parity between the irreducible subspaces~\footnote{
This unitary gate can be expressed as an exponential by considering $X=\sum_a T^a\otimes T^a$. $X$ then has two eigenvalues, $\lambda_{1}$ on the singlet and $\lambda_{\mathrm{Adj}}$ on the adjoint.  Defining
\begin{equation}
U=\exp\!\left[i\pi\,\frac{X-\lambda_{1}}{\lambda_{\mathrm{Adj}}-\lambda_{1}}\right]
\end{equation}
implements a relative $\pi$ phase between the two eigenspaces and hence reproduces $\textbf{U}=P_{1}-P_{\mathrm{Adj}}$ up to an overall phase.}
\begin{flalign}
	\mathbf{U}\ket{\psi}_{\textbf{1}}&= +\ket{\psi}_{\textbf{1}}\\
	\mathbf{U}\ket{\psi^{a}}_{\textbf{Adj}} &= - \ket{\psi^{a}}_{\textbf{Adj}}.
\end{flalign}

Let us for convenience refer to scattering on the $\textbf{N}\otimes \textbf{N}$ space as the $s$-channel, while $\textbf{N}\otimes \bar{\textbf{N}}$ as the $t$-channel, referring to the Mandelstam variables defined by $s=-(p_{1}+p_{2})^{2}, \quad\quad t = - (p_{1}-p_{3})^{2}$. The scattering amplitudes in either are expressed in their reducible form as
\begin{flalign}
\mathcal{M}^{(s)} =& \mathcal{M}(s,\Omega)\mathbf{S}_{\mathbb{I}}+ \mathcal{M}_{W}(s,\Omega)\mathbf{S}_{W},\\
\mathcal{M}^{(t)} =& \mathcal{M}^{\prime}(t,\Omega)\mathbf{S}_{\mathbb{I}} + \mathcal{M}_{U}(t,\Omega)\mathbf{U}.
\end{flalign}
Thus, any $SU(N)$-invariant 2-2 scattering amplitude between fundamental or antifundamental particles can be expressed entirely in terms of three unitaries $\mathbf{S}_{\mathbb{I}}$ , $\mathbf{S}_{W}$, and $\mathbf{U}$ up to scalar coefficients. We note that these coefficients are not independent and are related by crossing symmetry. Crossing imposes the relations among the quantum gates
\begin{equation}
	\begin{pmatrix}\mathbf{I}\\ \mathbf{S}_{W} \end{pmatrix} = \begin{pmatrix}\frac{N}{2} &\frac{N}{2}\\ 1 & 0 \end{pmatrix}\cdot \begin{pmatrix}\mathbf{I} \\\mathbf{U} \end{pmatrix}.
	\label{eq:crossing_gates}
\end{equation}
These relations may be viewed as the gate-level analogue of the standard recoupling coefficients appearing in field-theoretic treatments of crossing.  A detailed derivation of these relations will be presented elsewhere.
 
Remarkably, the two scattering channels in terms of quantum gates are captured by the single algebraic relation
\begin{equation}
\mathcal{M} = \mathcal{A}(s,t,u)\mathbf{S}_{\mathbb{I}} +  \mathcal{B}(s,t,u)\mathbb{Z},
\label{eq:algebra}
\end{equation}
where the set $\{\mathbf{S}_{\mathbb{I}}, \mathbb{Z}\}\simeq \mathbb{Z}_{2}$. On the $\textbf{N}\otimes \textbf{N}$ space $\mathbf{S}_{W}^{2} = \mathbf{S}_{\mathbb{I}} $, while on the $\textbf{N}\otimes \bar{\textbf{N}}$, $\mathbf{U}^{2}=\mathbf{S}_{\mathbb{I}}$. Note that this holds for any dimension of the qudit space, $N$.\\
\begin{figure}[t]
\centering
\begin{tikzpicture}[scale=1.15, thick]
  % ====== macros you can edit ======
  \def\x{0.75}   % s-channel point x  (coeff of S_I)
  \def\y{0.55}   % s-channel point y  (coeff of i Z_s)

  % Fill these after computing (a',b') via Eq. (35):
  % x' = Re(a'),  y' = Re(i b') = -Im(b')
  \def\xprime{1.10}   % <-- put Re(a') here
  \def\yprime{0.10}   % <-- put Re(i b') here

  % ====== background disk and axes ======
  \fill[gray!15] (0,0) circle (1.5);                          % all amplitudes
  \draw[gray!60, very thick] (0,0) circle (1.5);              % unitary locus
  \draw[->,gray!70] (-2.0,0) -- (2.0,0) node[right] {$\mathbf{S}_{\mathbb{I}}$};
  \draw[->,gray!70] (0,-2.0) -- (0,2.0) node[above] {$i\mathbb{Z}$};

  % ====== special boundary points ======
  \fill (1.5,0) circle (1.4pt) node[below right] {\scriptsize $\theta=0$ ($\mathcal{M}=\mathbf{S}_{\mathbb{I}}$)};
  \fill (-1.5,0) circle (1.4pt) node[below left] {\scriptsize $\theta=\tfrac{\pi}{2}$ ($\mathcal{M}\propto \mathbb{Z}$)};

  % ====== sample unitary point on boundary (kept from before) ======
  \fill ( {1.5*cos(38)}, {1.5*sin(38)} ) circle (1.4pt)
        node[above right] {\scriptsize $\mathcal{M}=e^{i\phi}(\cos\theta\,\mathbf{S}_{\mathbb{I}}+i\sin\theta\,\mathbb{Z})$};
  \draw[->,>=latex] ({1.5*cos(20)}, {1.5*sin(20)}) arc (20:38:1.5);
  \node at (0.75,0.55+0.22) {\scriptsize $\theta$};

  % ====== nonunitary s-channel point and its crossed image ======
  \fill[black] (\x,\y) circle (1.6pt) node[left] {\scriptsize $M=aS_I+bZ_s$};
  \fill[black] (\xprime,\yprime) circle (1.6pt) node[below left] {\scriptsize $\mathcal{M}'=a'\mathbf{S}_{\mathbb{I}}+b'\mathbb{Z}_t$};

  % Arrow showing crossing map
   \draw[-,thick,dotted] (\x,\y) -- (\xprime,\yprime);

  % Legend
  %\draw[gray!60] (1.75,-1.6) -- (2.0,-1.6);
  %\node[anchor=west] at (2.05,-1.6) {\scriptsize unitary locus};
  %\draw[gray!15, line width=3pt] (1.75,-1.9) -- (2.0,-1.9);
  %\node[anchor=west] at (2.05,-1.9) {\scriptsize all amplitudes};
\end{tikzpicture}
\caption{
Space of $SU(N)$-invariant 2-2 amplitudes in the 
two-operator subspace $\mathrm{span}\{\mathbf{S}_{\mathbb{I}},i\mathbb{Z}\}$. 
The shaded disk depicts the physically allowed region in any closed 
unitary sector (e.g.\ a fixed partial wave or helicity block); the boundary 
circle marks $|a_J|^2+|b_J |^2=1$. 
Coefficients $(a(s,\Omega),b(s,\Omega))$ for fixed $s$ may 
lie outside the disk prior to this projection. 
The dotted arrow illustrates a representative crossing transformation.}
\label{fig:amplitude-disk-crossing}
\end{figure}
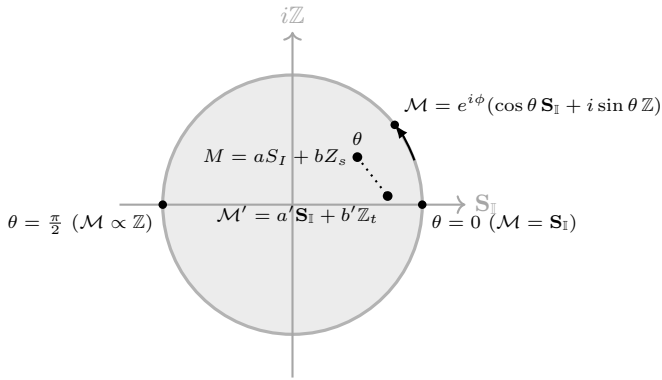
Eq.~\ref{eq:algebra} encapsulates the central result of this work: 
the space of possible $SU(N)$-invariant two-body amplitudes is entirely spanned by two involutive unitaries, $\{\mathbf{S}_{\mathbb{I}},\,\mathbb{Z}\}$, with $\mathbb{Z}\in\{\mathbf{S}_{W},\,\textbf{U}\}$.  
Every amplitude therefore lies in the two-dimensional operator subspace $\mathrm{span}\{\mathbf{S}_{\mathbb{I}},\mathbb{Z}\}$, remarkably mirroring the structure of a logical qubit.  
This reduction reveals that the apparent complexity of a continuous non-Abelian symmetry can be encoded in the minimal algebra generated by a pair of $\mathbb{Z}_2$ operations.

To see this more intuitively, consider
\begin{equation}
\mathcal{M}=a\mathbf{S}_{\mathbb{I}}+b\mathbb{Z},
\end{equation}
where $\mathcal{M}$ is unitary. Then
\begin{equation}
|a|^2+|b|^2=1,\qquad \mathrm{Re}(a^*b)=0.
\end{equation}
Parameterizing $a=e^{i\phi}\cos\theta$, $b=i\,e^{i\phi}\sin\theta$, we obtain
\begin{equation}
\mathcal{M}=e^{i\phi}\big(\cos\theta\mathbf{S}_{\mathbb{I}}+i\sin\theta\, \mathbb{Z}\big)
= e^{i\phi}\,e^{\,i\theta \mathbb{Z}},
\end{equation}
with eigenvalues $e^{i\phi}e^{\pm i\theta}$. Thus, in the invariant two–operator algebra $\mathrm{span}\{\mathbf{S}_{\mathbb{I}},\mathbb{Z}\}$, $\mathcal{M}$ is unitarily equivalent to a single–qubit $\mathbb{Z}$–rotation (up to a global phase). The geometry is summarized in Fig.~\ref{fig:amplitude-disk-crossing}: the shaded disk depicts the entire space of amplitudes defined by $(a,b)$, with the unitary cases on the boundary.~\footnote{We only show the case of $\mathcal{M}$ unitary for illustrative purposes. This is clearly not a realistic example connected to quantum field theories.}\\

It is useful to connect this picture with the partial-wave
decomposition introduced in ref.~\cite{McGinnis:2025brt}.  At fixed $s$ and helicities,
the 2-2 amplitude can be expanded in partial waves as
\begin{equation}
\langle mn|\mathcal{M}(s)|sr\rangle
  = \sum_J (2J+1)\,\big[a_J^{\{h\}}(s)\big]_{mn,sr}\,D^{J*}_{\{h\}}(\Omega)\,,
\end{equation}
where the Wigner $D$-functions, $D^{J}_{\{h\}}(\Omega)$, encode the
angular dependence.  Projecting onto a fixed $SU(N)$-invariant
pair of irreducible channels (e.g.\ $1\oplus\mathrm{Adj}$ or
$S\oplus A$), each partial wave $J$ defines a finite-dimensional
sector in which the internal $S$-matrix reduces to
\begin{equation}
\mathcal{M}_J(s) \;=\; a_J(s)\,\mathbf{S}_{\mathbb{I}} \;+\; b_J(s)\,\mathbb{Z}\,.
\end{equation}
Unitarity of the full S-matrix implies
that the eigenvalues of the $J$-th partial wave,
$S_{J,\pm}(s)=1 + i\,\kappa_J(s)\big(a_J\pm b_J\big)$, lie inside
the unit disk in the complex plane.  In the normalized convention
adopted here, this is equivalent to the bound
\begin{equation}
|a_J(s)|^2 + |b_J(s)|^2\;\le\; 1\,,
\end{equation}
with saturation corresponding to purely elastic scattering in that
partial-wave/helicity sector.  Thus, Fig.~\ref{fig:amplitude-disk-crossing}
can be interpreted as the space of allowed coefficients $(a_J,b_J)$
for any closed unitary sector, with each partial wave $J$ (and
choice of SU($N$) channel) represented by a point inside the disk. The relations in Eq.~(34) establish how the two channel bases of quantum gates are connected through $SU(N)$ recoupling via crossing symmetry. In Fig.~\ref{fig:amplitude-disk-crossing}, this relation between the two channels is shown schematically by the two points connected by the dotted line.

%%%%%%%%%%%%%%%%%%%%%%%%%%%%%%%%%%%%%%%%%%%%%%%%%%%%%%%%%%%%%%%%%%%%%%%%%%%%%%%%%%%%%%%%%%%%%%%%%%%%%%%%%%%%%%%%%%%%%%%%%%%%%

\noindent\textbf{\textit{Linear Combination of Unitaries and Block Encoding}} -   
Having identified the complete set of invariant operators and their transformation properties, it is now straightforward to examine how these amplitudes can be represented and implemented in the language of quantum computation.  
In particular, the operators forming each channel can be expressed as linear combinations of unitaries~\cite{Childs:2012gwh,Berry:2014ivo,BerryChildsKothari2015}, which naturally admit a block-encoded circuit realization~\cite{GilyenEtAl2019,LowChuang2019}.

The operator form of the $SU(N)$-invariant scattering amplitude,
\begin{equation}
\mathcal{M} = a\,\mathbf{S}_{\mathbb{I}} + b\,\mathbb{Z} , 
\qquad \mathbb{Z}\in\{\mathbf{S}_{W},\,\textbf{U}\},
\label{eq:LCU_form}
\end{equation}
admits a direct realization as a \textit{linear combination of unitaries} (LCU).  
The coefficients $a,b$ are complex scalars encoding the kinematic dependence of the amplitude.  
By absorbing the phases of $a$ and $b$ into redefined unitaries
\begin{equation}
U_I = e^{i\phi_a}\mathbf{S}_{\mathbb{I}} , 
\qquad 
U_Z = e^{i\phi_b}\mathbb{Z} ,
\end{equation}
where $a=|a|e^{i\phi_a}$ and $b=|b|e^{i\phi_b}$, the operator can be expressed in terms of positive real weights $|a|$ and $|b|$.  
Let us define the normalization and mixing angle
\begin{equation}
\alpha = |a| + |b|, 
\qquad 
\cos\gamma = \sqrt{\frac{|a|}{\alpha}}, 
\qquad 
\sin\gamma = \sqrt{\frac{|b|}{\alpha}} .
\end{equation}
Then the amplitude $\mathcal{M}/\alpha$ can be implemented as a \textit{one-ancilla block encoding}, using the unitary
\begin{flalign}\nonumber
W = R_y&(-2\gamma)\otimes I\\
      &\times\Big[(|0\rangle\!\langle0|\otimes U_I)
          +(|1\rangle\!\langle1|\otimes U_Z)\Big]
      R_y(2\gamma)\otimes I,
\label{eq:block_encoding_unitary}
\end{flalign}
which satisfies the block-encoding identity
\begin{equation}
(\langle 0|\!\otimes\! I)\, W\, (|0\rangle\!\otimes\! I)
= \frac{\mathcal{M}}{\alpha}.
\label{eq:block_encoding_identity}
\end{equation}
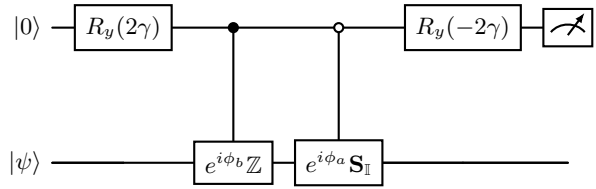
\begin{figure}[t]
\begin{quantikz}[row sep=1.2cm, column sep=0.3cm]
\lstick{$\ket{0}$} 
    & \gate{R_y(2\gamma)} 
    & \ctrl{1} 
    & \octrl{1} 
    & \gate{R_y(-2\gamma)} 
    & \meter{} \\
\lstick{$\ket{\psi}$}
    & \qw 
    & \gate{e^{i\phi_b} \mathbb{Z}} 
    & \gate{e^{i\phi_a} \mathbf{S}_{\mathbb{I}}} 
    & \qw 
    & \qw
\end{quantikz}
\caption{Block encoding of 2-2 $SU(N)$-invariant scattering amplitudes.}
\label{fig:circuit}
\end{figure}
Operationally, the rotation $R_y(2\gamma)$ prepares the ancilla state
$\sqrt{|a|/\alpha}\ket{0} + \sqrt{|b|/\alpha}\ket{1}$; 
the controlled gates apply $U_I$ or $U_Z$ on the system depending on the ancilla value; 
and the final inverse rotation unprepares the superposition.  
The total circuit $W$ acts unitarily on the combined ancilla–system space, and 
either postselection on the ancilla or \textit{oblivious amplitude amplification} can be used to implement $\mathcal{M}$ coherently within a larger computation.  
The construction requires only one ancilla qubit and two controlled gates, 
independent of $N$. We show the explicit circuit construction in Fig.~\ref{fig:circuit}.

For equal-weight coefficients ($|a|=|b|$), 
the rotation reduces to a Hadamard gate $H$ in place of $R_y(2\gamma)$, reproducing the canonical LCU circuit.  
Because $\mathbf{S}_{\mathbb{I}}$, $\mathbf{S}_{W}$, and $\textbf{U}$ are each unitary reflections, the same circuit applies in either scattering channel with $\mathbb{Z}=\mathbf{S}_{W}$ or $\mathbb{Z}=\textbf{U}$.  
The overhead is constant with respect to the qudit dimension $N$; 
all $N$-dependence resides in the internal implementation of the two-particle gates $\mathbf{S}_{W}$ and $\textbf{U}$.

This construction shows that every $SU(N)$-invariant 2-2 amplitude between particles transforming in the fundamental or antifundamental representations is 
\textit{block-encodable} with a single ancilla qubit.  
The $S$-matrix thus admits a concrete realization as a quantum circuit built from 
involutive unitaries, providing a direct bridge between the algebraic formulation of global symmetries and scattering theory and the language of quantum computation.\\

%%%%%%%%%%%%%%%%%%%%%%%%%%%%%%%%%%%%%%%%%%%%%%%%%%%%%%%%%%%%%%%%%%%%%%%%%%%%%%%%%%%%%%%%%%%%%%%%%%%%%%%%%%%%%%%%%%%%%%%%%%%%%%%%%%%%%%%%%%%%%%%%%%%%%%%%%%%%%%%%%%%%%%%%%%%%%%%%
%\section{\label{sec:conclusion}Conclusions}

\noindent\textbf{\textit{Conclusions}} - In this work we have shown that all $SU(N)$-invariant 2-2 scattering amplitudes between fundamental or antifundamental particles, i.e. qudits, can be constructed from a minimal set of three unitary gates, $\{\mathbf{S}_{\mathbb{I}},\,\mathbf{S}_{W},\,\textbf{U}\}$, each possible scattering channel forming a $Z_2$ algebra.  This result identifies the scattering amplitudes as quantum circuits whose logical structure is generated entirely by involutive unitaries, providing an unexpectedly simple computational substrate for continuous non-Abelian symmetries.  The entire group structure of $SU(N)$ emerges from the composition of these discrete reflections, demonstrating that symmetry in this context can be viewed as an algebraic extension of binary gate operations.

The block-encoding construction makes this correspondence operational.  The scattering amplitude can be implemented as a linear combination of unitaries controlled by a single ancilla qubit, revealing that the information contained in scattering amplitudes is efficiently representable within standard quantum algorithmic primitives.  This not only connects scattering theory with the framework of quantum simulation, but also suggests a route to classifying interactions by their computational complexity, independent of any particular dynamical model.

More broadly, these findings indicate that global symmetries and quantum logic are not independent notions.  The correspondence uncovered here raises several questions for future work, including implementation of amplitudes beyond the two-body sector, scattering where particle number is no longer conserved, e.g. $2\to3$ processes, and implementation of other groups and representations.  In this sense, the computational structure of $SU(N)$ scattering may serve as a prototype for a more general correspondence between the logical operations of quantum computation and the invariant operators of quantum field theory.

In summary, we have shown that $SU(N)$-invariant 2-2 scattering can be realized as a linear combination of three involutive unitaries forming a $\mathbb{Z}_{2}$ algebra.  This identifies scattering amplitudes as a minimal quantum circuit whose discrete computational structure encodes continuous global symmetries realized in quantum field theories.  Extensions to higher-point amplitudes, mixed representations, and gauge symmetries may reveal further connections between the logical and group-theoretic foundations of quantum field theory.\\

%%%%%%%%%%%%%%%%%%%%%%%%%%%%%%%%%%%%%%%%%%%%%%%%%%%%%%%%%%%%%%%%%%%%%%%%%%%%%%%%%%%%%%%%%%%%%%%%%%%%%%%%%%%%%%%%%%%%%%%%%%%%%%%%%%
\begin{acknowledgments}
This work has been supported in part by the U.S. Department of Energy under grant No. DEFG02-13ER41976/DE-SC0009913.
\end{acknowledgments}

%\appendix

\nocite{*}

\bibliography{apssamp}% Produces the bibliography via BibTeX.

% If you want this specific reference and are using BibTeX,
% put the entry in apssamp.bib instead of a manual \bibitem here.
% \bibitem{BerryChildsKothari2015}
% D. W. Berry, A. M. Childs, and R. Kothari,
% ``Hamiltonian Simulation with Nearly Optimal Dependence on all Parameters,''
% in {\it Proceedings of the 2015 IEEE Annual Symposium on Foundations of Computer Science},
% pp. 792--809 (2015), doi:10.1109/FOCS.2015.54.

\end{document}